# Analysis of a Fivefold Symmetric Superposition of Plane Waves

Michael H. Schwarz

Department of Mathematical Sciences, Rensselaer Polytechnic Institute, Troy, New York 12180, USA

Robert A. Pelcovits

Department of Physics, Brown University, Providence, Rhode Island 02912, USA

We show that a symmetric superposition of five standing plane waves can be expressed as an infinite series of terms of decreasing wavenumber, where each term is a product of five plane waves. We show that this series converges pointwise in $\mathbb{R}^2$ and uniformly in any disk domain in $\mathbb{R}^2$. Using this series, we provide a heuristic argument for why the locations of the local extrema of a symmetric superposition of five standing plane waves can be approximated by the vertices of a Penrose tiling.

In a previous paper [1], we found that the vertices of a Penrose tiling approximate the local extrema of a fivefold symmetric superposition of standing plane waves. In this paper, we show that such a superposition of plane waves is equal to an infinite series of products of plane waves. This equality factors into a heuristic argument for why the Penrose tiling approximation of the local extrema is effective.

Ignoring time dependence, the fivefold symmetric superposition of standing plane waves that is of interest is given by:

$$s_{5,k}(x, y) \equiv \sum_{i=0}^{4} \sin(kA_i(x, y))$$

where k is the wavenumber, and

$$A_i(x, y) = (x, y) \bullet \vec{e}_i, \text{ where } \vec{e}_i = \left(\cos\left(\frac{2\pi i}{5}\right), \sin\left(\frac{2\pi i}{5}\right)\right)$$

The motivation for trying to describe $s_{5,k}(x, y)$ in terms of a product or multiple products of plane waves comes from a simpler case, a superposition of two standing plane waves that are perpendicular to each other, given by

$$s_{2,k}(x, y) \equiv \sin(kx) + \sin(ky).$$

In this case, the following equality holds:

$$\sin(kx) + \sin(ky) = 2\sin\left(\frac{k[x+y]}{2}\right)\cos\left(\frac{k[x-y]}{2}\right)$$

The expression $2\sin\left(\frac{k[x+y]}{2}\right)\cos\left(\frac{k[x-y]}{2}\right)$ is equal to 0 whenever $\sin\left(\frac{k[x+y]}{2}\right) = 0$ or $\cos\left(\frac{k[x-y]}{2}\right) = 0$, yielding a square grid in $\mathbb{R}^2$ on which $s_{2,k}(x,y) = 0$. This grid divides $\mathbb{R}^2$ into a "checkerboard", half of whose squares have $s_{2,k}(x,y) < 0$, with $s_{2,k}(x,y) > 0$ in the other squares. Each square contains one local maximum or one local minimum of $s_{2,k}(x,y)$. These local extrema are located at the centers of the squares and thus coincide with the vertices of the dual of the aforementioned grid.

Given the effectiveness of this approach in describing the local extrema of a superposition of two standing waves, the question arises of whether a similar approach works in the fivefold symmetric case. It turns out that $s_{5,k}(x,y)$ cannot be described as a product of five plane waves, but it can be described as a series of such products. The series, which is the main result of this paper, is as follows:

$$\sum_{i=0}^{4} \sin(kA_i(x,y)) = 16 \sum_{n=0}^{\infty} (-1)^n F_n \prod_{i=0}^{4} \sin\left(\frac{k}{2\tau^{n+1}} A_i(x,y)\right)$$

where $A_i(x,y) \equiv (x,y) \bullet \vec{e}_i$,

$$\vec{e}_i \equiv \left(\cos\left(\frac{2\pi i}{5}\right), \sin\left(\frac{2\pi i}{5}\right)\right),$$

$$\tau \equiv \frac{1+\sqrt{5}}{2}, \text{ and}$$

$F_n$ is the nth Fibonacci number, i.e. $F_0 = 1, F_1 = 1, F_2 = 2, F_3 = 3, F_4 = 5, etc.$

This result is stated in terms of earlier definitions as

$$s_{5,k}(x,y) = 16 \sum_{n=0}^{\infty} (-1)^n F_n p_{5,k/(2\tau^{\wedge}[n+1])}(x,y)$$

where $p_{5,k}(x,y) \equiv \prod_{i=0}^{4} \sin(kA_i(x,y))$, and $A_i, \vec{e}_i, \tau$, and $F_n$ are defined as above.

The proof of this result is as follows:

From $\exp(jk[A_0 + A_1 + A_2 + A_3 + A_4]) = \exp(jkA_0)\exp(jkA_1)\exp(jkA_2)\exp(jkA_3)\exp(jkA_4)$, where $j \equiv \sqrt{-1}$, it follows that:

$$\sin(k[A_0 + A_1 + A_2 + A_3 + A_4])$$
$$-\begin{bmatrix}\sin(k[-A_0 + A_1 + A_2 + A_3 + A_4]) + \sin(k[A_0 - A_1 + A_2 + A_3 + A_4]) + \sin(k[A_0 + A_1 - A_2 + A_3 + A_4]) \\ + \sin(k[A_0 + A_1 + A_2 - A_3 + A_4]) + \sin(k[A_0 + A_1 + A_2 + A_3 - A_4])\end{bmatrix}$$
$$+\begin{bmatrix}\sin(k[-A_0 - A_1 + A_2 + A_3 + A_4]) + \sin(k[A_0 - A_1 - A_2 + A_3 + A_4]) + \sin(k[A_0 + A_1 - A_2 - A_3 + A_4]) \\ + \sin(k[A_0 + A_1 + A_2 - A_3 - A_4]) + \sin(k[-A_0 + A_1 + A_2 + A_3 - A_4])\end{bmatrix}$$
$$+\begin{bmatrix}\sin(k[-A_0 + A_1 - A_2 + A_3 + A_4]) + \sin(k[A_0 - A_1 + A_2 - A_3 + A_4]) + \sin(k[A_0 + A_1 - A_2 + A_3 - A_4]) \\ + \sin(k[-A_0 + A_1 + A_2 - A_3 + A_4]) + \sin(k[A_0 - A_1 + A_2 + A_3 - A_4])\end{bmatrix}$$
$$= 16\sin(kA_0)\sin(kA_1)\sin(kA_2)\sin(kA_3)\sin(kA_4)$$

Since $A_0 + A_1 + A_2 + A_3 + A_4 = 0$, $A_0 + A_1 = -\tau A_3$, and $A_0 + A_2 = \frac{1}{\tau}A_1$, and since cyclic permutations of these relationships hold as well (e.g. $A_1 + A_2 = -\tau A_4$), the above simplifies to:

$$\sin(0)$$
$$-[\sin(-2kA_0) + \sin(-2kA_1) + \sin(-2kA_2) + \sin(-2kA_3) + \sin(-2kA_4)]$$
$$+[\sin(2k\tau A_3) + \sin(2k\tau A_4) + \sin(2k\tau A_0) + \sin(2k\tau A_1) + \sin(2k\tau A_2)]$$
$$+\left[\sin\left(-\frac{2k}{\tau}A_1\right) + \sin\left(-\frac{2k}{\tau}A_2\right) + \sin\left(-\frac{2k}{\tau}A_3\right) + \sin\left(-\frac{2k}{\tau}A_4\right) + \sin\left(-\frac{2k}{\tau}A_0\right)\right]$$
$$= 16\sin(kA_0)\sin(kA_1)\sin(kA_2)\sin(kA_3)\sin(kA_4)$$

This is expressed in terms of earlier definitions as:

$$s_{5,2k} + s_{5,2k\tau} - s_{5,2k/\tau} = 16 p_{5,k}$$

This relationship is satisfied by:

$$s_{5,k}(x,y) = 16 \sum_{n=0}^{\infty} (-1)^n F_n p_{5,k/(2\tau^\wedge[n+1])}(x,y)$$

provided that the series converges pointwise in $R^2$. The series does converge pointwise in $R^2$ and also converges uniformly in any disk domain in $R^2$. This convergence is proven as follows:

Let $r = \|(x, y)\|$

$|\sin(kA_i(x, y))| \leq kr \; \forall \; i \in \{0,1,2,3,4\}$

$|p_{5,k}(x, y)| = \left|\prod_{i=0}^{4} \sin(kA_i(x, y))\right| = \prod_{i=0}^{4} |\sin(kA_i(x, y))| \leq (kr)^5$

For any nonnegative integer N:

$$\left|\sum_{n=N}^{\infty} (-1)^n F_n p_{5,k/(2\tau^{\wedge}[n+1])}(x, y)\right| \leq \sum_{n=N}^{\infty} \left|(-1)^n F_n p_{5,k/(2\tau^{\wedge}[n+1])}(x, y)\right| = \sum_{n=N}^{\infty} F_n \left|p_{5,k/(2\tau^{\wedge}[n+1])}(x, y)\right|$$

$$\left|\sum_{n=N}^{\infty} (-1)^n F_n p_{5,k/(2\tau^{\wedge}[n+1])}(x, y)\right| \leq \sum_{n=N}^{\infty} F_n \left(\frac{kr}{2\tau^{n+1}}\right)^5 = \sum_{n=N}^{\infty} \frac{\tau^{n+1} - \left(-\frac{1}{\tau}\right)^{n+1}}{\sqrt{5}} \left(\frac{kr}{2\tau^{n+1}}\right)^5$$

$$\left|\sum_{n=N}^{\infty} (-1)^n F_n p_{5,k/(2\tau^{\wedge}[n+1])}(x, y)\right| \leq \sum_{n=N}^{\infty} \frac{\tau^{n+1} + 1}{\sqrt{5}} \left(\frac{kr}{2\tau^{n+1}}\right)^5 = \frac{1}{\sqrt{5}} \sum_{n=N}^{\infty} \tau^{n+1} \left(\frac{kr}{2\tau^{n+1}}\right)^5 + \frac{1}{\sqrt{5}} \sum_{n=N}^{\infty} \left(\frac{kr}{2\tau^{n+1}}\right)^5$$

$$\left|\sum_{n=N}^{\infty} (-1)^n F_n p_{5,k/(2\tau^{\wedge}[n+1])}(x, y)\right| \leq \frac{1}{\sqrt{5}} \left(\frac{kr}{2}\right)^5 \left(\sum_{n=N}^{\infty} \left(\frac{1}{\tau^{n+1}}\right)^4 + \sum_{n=N}^{\infty} \left(\frac{1}{\tau^{n+1}}\right)^5\right)$$

$$\left|\sum_{n=N}^{\infty} (-1)^n F_n p_{5,k/(2\tau^{\wedge}[n+1])}(x, y)\right| \leq \frac{1}{\sqrt{5}} \left(\frac{kr}{2}\right)^5 \left(\frac{1}{\tau^{4N+4}} \frac{1}{1 - \left(\frac{1}{\tau^4}\right)} + \frac{1}{\tau^{5N+5}} \frac{1}{1 - \left(\frac{1}{\tau^5}\right)}\right)$$

For any disk domain of radius R centered on the origin, $\frac{1}{\sqrt{5}} \left(\frac{kr}{2}\right)^5$ has an upper bound C given

by $C = \frac{1}{\sqrt{5}} \left(\frac{kR}{2}\right)^5$. Consequently, for (x,y) in such a disk:

$$\left|\sum_{n=N}^{\infty} (-1)^n F_n p_{5,k/(2\tau^{\wedge}[n+1])}(x, y)\right| \leq C \left(\frac{1}{\tau^{4N+4}} \frac{1}{1 - \left(\frac{1}{\tau^4}\right)} + \frac{1}{\tau^{5N+5}} \frac{1}{1 - \left(\frac{1}{\tau^5}\right)}\right)$$

The limit of the right side of this inequality as N goes to infinity is 0, and is independent of (x,y), proving uniform convergence of the series $\sum_{n=0}^{\infty} (-1)^n F_n p_{5,k/(2\tau^{\wedge}[n+1])}(x, y)$ for any disk domain centered on the origin. It follows that the series converges uniformly in any disk domain in $R^2$, and pointwise in $R^2$.

This completes the proof that $s_{5,k}(x, y) = 16 \sum_{n=0}^{\infty} (-1)^n F_n p_{5,k/(2\tau^{[n+1]})}(x, y)$.

A heuristic argument for why a Penrose tiling can be used to approximate the locations of the local extrema of $s_{5,k}(x, y)$ follows from this equality. The argument is as follows. The locations of the local extrema of a series of oscillating terms are often well-approximated by the locations of the local extrema of the highest wavenumber term in the series. The highest wavenumber term of the series above is the first term, $\sum_{n=0}^{\infty} p_{5,k/(2\tau)}(x, y)$, so a reasonable conjecture is that the local extrema of this term have approximately the same locations as those of $s_{5,k}(x, y)$. The set of points satisfying $\sum_{n=0}^{\infty} p_{5,k/(2\tau)}(x, y) = 0$ is a set of five sets of parallel lines, known as a pentagrid, whose dual is a Penrose tiling. The pentagrid divides $R^2$ into regions where $\sum_{n=0}^{\infty} p_{5,k/(2\tau)}(x, y)$ is positive and regions were $\sum_{n=0}^{\infty} p_{5,k/(2\tau)}(x, y)$ is negative. Each of these regions must contain at least one local extremum of $\sum_{n=0}^{\infty} p_{5,k/(2\tau)}(x, y)$ since $\sum_{n=0}^{\infty} p_{5,k/(2\tau)}(x, y) = 0$ on the boundary of such a region. Given that $s_{5,k}(x, y)$ does not oscillate too rapidly relative to the size of these regions, it is reasonable to expect that there is at most one local extremum in each region. It is also reasonable to expect that this local extremum occurs roughly at the center of this region. Since the dual of the pentagrid is a Penrose tiling, the vertices of the Penrose tiling are generally expected to be near the centers of these regions, and consequently near the local extrema in these regions. This leads to the expectation that the local extrema of $s_{5,k}(x, y)$ are near the vertices of the Penrose tiling.

In conclusion, we have shown that a symmetric superposition of five standing plane waves can be expressed as an infinite series of products of five plane waves. This series converges pointwise in $R^2$ and uniformly in any disk domain in $R^2$. The series is useful in a heuristic argument for why a Penrose tiling can be used to approximate the locations of the local extrema of a symmetric superposition of five standing plane waves.